\documentstyle[aasms4,psfig]{article}

\def\ion#1#2{#1$\;${\small\rm{#2}}\relax}
\def\ltsima{$\; \buildrel < \over \sim \;$}
\def\simlt{\lower.5ex\hbox{\ltsima}}
\def\kms{{\rm\,km\,s^{-1}}}
\def\kpc{{\rm\,kpc}}
\def\etal{{\it et al.\ }}
\newcommand{\angstrom}{\mbox{\textrm{\AA}}}
\newcommand\comment[1]{\null}

\begin{document}
\title{An Investigation of Gravitational Lensing in the \\ Southern 
BL Lac PKS 0537-441}
\slugcomment{Accepted for Publication by the Astrophysical Journal}

\lefthead{Lewis \& Ibata}
\righthead{PKS 0537-441}

\author{
Geraint F. Lewis\altaffilmark{1} \& Rodrigo A. Ibata\altaffilmark{2}
\altaffiltext{1}{ 
Astronomy Dept., University of Washington, Seattle WA, U.S.A.
\& \\ 
Dept. of Physics and Astronomy, University of Victoria, 
Victoria, B.C., Canada\\
Electronic mail: {\tt gfl@astro.washington.edu} \\
Electronic mail: {\tt gfl@uvastro.phys.uvic.ca}
}
\altaffiltext{2}
{European Southern Observatory, Garching bei M\"unchen, Germany \\
Electronic mail: {\tt ribata@eso.org}}
}

\begin{abstract}
The  BL--Lac  family of  active  galaxies  possess almost  featureless
spectra  and  exhibit rapid  variability  over  their entire  spectral
range.   A number  of  models  have been  developed  to explain  these
extreme  properties,  several of  which  have  invoked  the action  of
microlensing by sub--stellar mass objects in a foreground galaxy; this
not  only  introduces variability,  but  also  amplifies an  otherwise
normal  quasar  source.   Here  we  present  recent  spectroscopy  and
photometry  of  the southern  BL~Lac  PKS~0537-441;  with an  inferred
redshift of $z\sim0.9$ it represents  one of the most distant and most
luminous members of  the BL~Lac family.  The goal  of the observations
was  not only to  confirm the  redshift of  PKS~0537-441, but  also to
determine the redshift of a putative galaxy along the line of sight to
the BL--Lac;  it has  been proposed  that this galaxy  is the  host of
microlensing stars that account for PKS~0537-441's extreme properties.
While several observations have failed to detect any extended emission
in  PKS~0537-441, the  HST imaging  data presented  here  indicate the
presence of  a galactic  component, although we  fail to  identify any
absorption features that  reveal the redshift of the  emission.  It is
also  noted that  PKS~0537-441 is  accompanied by  several  small, but
extended  companions, located  a few  arcseconds from  the point--like
BL--Lac  source.  Two  possibilities present  themselves;  either they
represent  true   companions  of  PKS~0537--441,   or  are  themselves
gravitationally lensed images of more distant sources.
\end{abstract}

\newcommand{\pk}{PKS~0537--441}
\newcommand{\bl}{BL--Lac}
\newcommand{\bls}{BL--Lacs}

\newcommand{\ffffff}[1]{\mbox{$#1$}}
\newcommand{\scnd}{\mbox{\ffffff{''}\hskip-0.3em.}}
\newcommand{\scmd}{\mbox{\ffffff{''}}}

\section{Introduction}
\bl\ systems represent an extreme  class of active galaxy that exhibit
featureless continua and undergo  rapid variability at all wavelengths
(c.f. Urry et  al.  1997).  A variety of models  have been proposed to
explain this extreme behavior, including both intrinsic processes such
as  relativistic jet accelerations  (Georganopoulos \&  Marscher 1998)
and  external effects  such as  interstellar scintillation  (Wagner \&
Witzel 1995).  Another  popular model proposes that \bls\  are in fact
Optically  Violent Variable  (OVV)  quasars whose  continua have  been
amplified by  microlenses in a  foreground galaxy (Ostriker  \& Vietri
1985); here  enhanced  continuum  swamps  the quasar  emission  lines,
resulting  in  the  featureless  spectrum  which  characterizes  \bls.
However,  while this  model explains  the spectral  characteristics of
\bls\  the  source  of  the  variability is  still  considered  to  be
intrinsic  to the  OVV source.   More  recent variants  of this  model
propose  the microlensed  source  is a  relativistic  jet, leading  to
extremely rapid variability (Gopal-Krishna \& Subramanian 1991), while
Nottale (1986) developed a  microlensing model utilizing very low mass
stars to account  for both the variability and  spectral properties of
\bl\ systems.

Although  problems  exist with  this  microlensing hypothesis  (Kayser
1988; Gear 1991; Lewis \&  Williams 1997), observationally a number of
\bl\  systems  appear  to  be superimposed  on  extended  nebulosities
(Stickel \etal\ 1988).   While it is obvious that  in several of these
the  nebulosity and  the  \bl\ source  are  physically associated,  in
others  the  extended emission  appears  to  be  a foreground  galaxy,
distinct  from the \bl\  source (Stocke  \etal\ 1985),  bolstering the
case for the  microlensing of \bls.  For microlensing  to be effective
at  introducing  rapid variability,  a  substantial  optical depth  of
stellar objects is  required and the \bl\ source  must possess a small
impact parameter  with respect to  the foreground galaxy.   With this,
however, macrolensing effects become important and most BL-Lacs should
appear offset  from the core of  any foreground lensing  galaxy.  In a
fraction of systems, the impact parameter should be sufficiently small
that the  source will  lie within the  multiple imaging region  of the
lens, leading to several images of the same \bl\ source about the core
of  the galaxy  (Narayan  \& Schneider  1990).  Observational  studies
reveal  {\bl}s to be  well-centered upon  their hosts  (Abraham \etal\
1991) and  it has  been proposed  that  the redshift  of several  \bl\
systems have  been mis-identified  (Narayan \& Schneider  1990); given
the  generally   featureless  spectrum  exhibited  by   \bls\  such  a
conclusion is very likely.

The purpose of  this paper is to explore  a particular bright, distant
\bl\  system,  \pk,  with  the   aim  of  understanding  the  role  of
microlensing in explaining its observed properties with a confirmation
its  redshift, as  well as  determining the  redshift of  the putative
galaxy  along  the line-of-sight.   The  layout  of  the paper  is  as
follows; Section~\ref{pks}  details the previous  observations of \pk\
and  the  historical  development  of  the  microlensing  model.   New
observations  of \pk\  are discussed  in  Section~\ref{obs}, detailing
both the NTT spectroscopy and HST imaging presented in this paper.  In
Section~\ref{conclusions} the conclusions of this study are presented.

\section{\pk}\label{pks}
Selected as a quasi-stellar source  in the Parkes 2700MHz radio source
(Peterson  \etal\ 1976),  \pk~is seen  to be  a bright  $(B  \sim 15)$
point--like  source.  A  search for  \pk\ on  the  Harvard Observatory
photographic plate collection  (Liller 1974) allowed the determination
of a light  curve of almost a century  in extent, revealing long--term
variations of  $\sim$5 mag with  fluctuations of $\sim$2  mag occuring
over  less  than two  months.   While  \bl\  systems normally  exhibit
featureless  spectra, weak  emission lines  are sometimes  seen during
quiescent periods.  This  was found to be the case  for \pk\ which was
observed to possess a  prominent emission feature at 5304\angstrom; it
was proposed that this line is Mg II $\lambda2798$, placing \bl\ at $z
= 0.894$ (Peterson \etal 1976; Stickel, Fried \& Kuehr 1993).  At this
redshift the  inferred properties of  \pk\ places it amongst  the most
luminous of blazars.  It was  also suggested that C III] $\lambda1909$
was  also seen  in these  early spectra,  although this  has  not been
confirmed in subsequent observations.

Imaging by Stickel \etal\  (1988) revealed that \pk\ lies superimposed
within  a system of  low redshift  galaxies.  One  of these,  a spiral
galaxy at  $z =  0.186$, is seen  $11$\scmd\ eastward, while  \pk\ was
inferred to lie  at the core of a galaxy  with an extended exponential
disk profile.  However, deeper  follow-up exposures in the optical and
NIR with the NTT and ESO  2.2m telescope failed to detect any extended
emission  around \pk\  (Falomo \etal\  1992; Kotilainen  \etal\ 1998),
suggesting that the findings of Stickel \etal\ (1988) were erroneous.

Given  its  apparent  brightness,  \pk\  is an  excellent  target  for
studying the properties of \bl\  objects and it has therefore been the
subject  of  several   monitoring  campaigns  at  various  wave-bands.
Recently it was  the subject of a monitoring campaign  over a two week
period at radio wavelengths (Romero \etal\ 1994), which revealed large
variations, up  to a factor of  $\sim 1.7$ on time-scales  as short as
$\sim 10^4$ secs.  Romero~\etal\ (1995) interpreted these as being due
to  microlensing  of  a  relativistic  shock--front by  stars  in  the
putative foreground galaxy.  The  duration of the events $\simlt 1.2$~
hours implies  that, if  the variability is  microlensing-induced, the
lenses must consist of sub-solar mass stars in the range $\sim 10^{-4}
- 10^{-3} {\rm M_{\odot}}$.

\section{Observations of \pk}\label{obs}

\subsection{Spectroscopy}

\subsubsection{Observations}
The spectroscopic  observations were  obtained on the  red arm  of the
EMMI instrument at the NTT, operating in service mode.  Six exposures,
of   total  duration   6600~sec  ($3\times   1000$~sec   and  $3\times
1200$~sec), were obtained by observatory staff on Nov 8, 1997, and Jan
3 and 5,  1998.  Taken with the Tek2k CCD  ($24\mu$m pixels) and grism
\#4,  the spectra  cover the  wavelength range  $5800$--$11000$\AA\ at
2.8\AA/pixel  resolution.  The gain  was set  to 1.1  ${\rm e^-/ADU}$,
giving a read-noise of ${\rm  4.2 e^-}$ RMS, and the spectrograph slit
was oriented North-South.  The seeing at the time of the observations,
determined from stellar profiles  seeing in short pointing images, was
better than $1\scmd$ for all the exposures.

Though  the grism is  reasonably efficient  over the  wavelength range
covered by our  spectra (more than 60\% efficient  redward of 7000\AA,
but dropping to 25\% at 5800\AA), the system response drops rapidly at
the red end of our spectral range due to the rapid drop in the quantum
efficiency of the CCD ($\sim 8$\% efficient at 10000\AA).

\subsubsection{Processing}
The  spectra  were  pre-processed  in the  usual  fashion  (debiasing,
flat-fielding).   So as  to  establish whether  any extended  emission
could  be seen in  the spectra,  for each  exposure, we  extracted the
spectrum perpendicular  to the dispersion  direction in four  bands of
width  $1\scmd$,  $2\scmd$,   $4\scmd$,  and  $6\scmd$.   Sky-emission
features were subtracted at this stage using the observed sky spectrum
$10\scmd$ away from the blazar as a template.  These spectra were then
wavelength calibrated  and observations  of the standard  star LTT2415
were  used to  remove  the telluric  features  and flux-calibrate  the
spectra.

\subsubsection{Results}
\begin{figure*}
\vbox{
\centerline{
\psfig{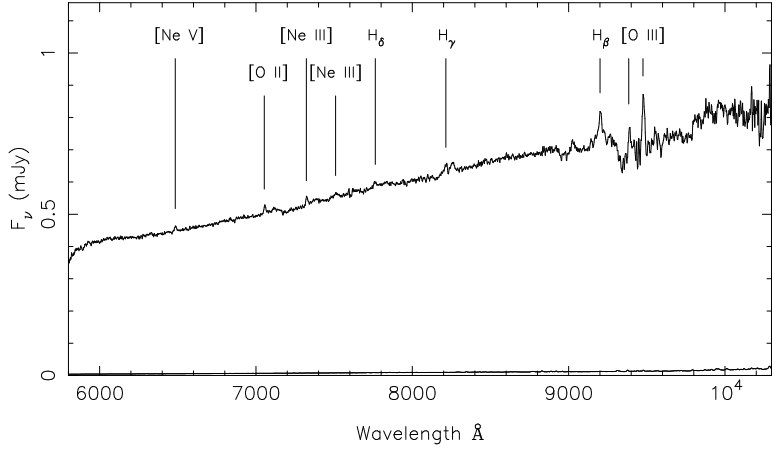}
}
\caption[]{The co-added spectrum of  \pk\ from the central $1\scmd$ of
the   NTT  observations.   Several   emission  features   are  visible
including;  [\ion{Ne}{V}]$\lambda 3426$,  [\ion{O}{II}]$\lambda 3727$,
[\ion{Ne}{III}]$\lambda                 3869$+\ion{He}{I}$\lambda3889$,
[\ion{Ne}{III}]$\lambda          3968$,          [\ion{S}{II}]$\lambda
4068/4076$+H$_\delta              \lambda4102$,              H$_\gamma
\lambda4340$+[\ion{O}{III}]$\lambda4363$,   H$_\beta  \lambda4861$  \&
[\ion{O}{III}]$\lambda  4959/5007$.  Beyond  9000\AA\ the  spectrum is
affected  by spurious  effects  from the  correction  of the  telluric
absorption features.  The noise spectrum is also presented.}
\label{fig1}}
\end{figure*}

Figure~\ref{fig1}   presents   the   co-added   spectrum   (with   the
corresponding  noise spectrum)  in the  $1\scmd$ extracted  band.  The
flux in the  $6\scmd$ extracted spectrum is a  factor of $\sim6$ times
lower than  that presented  by Stickel~\etal\ (1988),  indicating that
\pk\ was in  a quiescent state.  Our spectrum is  also flatter, with a
slope of $\alpha=-1.3$  (${\rm F_\nu\propto \nu^\alpha}$), in contrast
to -1.6 seen  by Stickel~\etal\ (1988). This low  state of activity in
the \bl\ proved to be very  fortuitous for this project as a number of
emission features are  visible including; [\ion{Ne}{V}]$\lambda 3426$,
[\ion{O}{II}]$\lambda          3727$,          [\ion{Ne}{III}]$\lambda
3869$+\ion{He}{I}$\lambda3889$,     [\ion{Ne}{III}]$\lambda     3968$,
[\ion{S}{II}]$\lambda   4068/4076$+H$_\delta  \lambda4102$,  H$_\gamma
\lambda4340$+[\ion{O}{III}]$\lambda4363$,   H$_\beta  \lambda4861$  \&
[\ion{O}{III}]$\lambda   4959/5007$   (c.f.   Fancis   \etal\   1991).
Combining the centroid positions of these lines, we determine that the
redshift of the \bl\ source  in \pk\ is ${\rm z=0.892\pm0.001}$, which
confirms earlier measurements.

However, note  that no absorption features indicative  of the presence
of an additional galaxy spectrum  are present.  In particular, we find
no evidence  for the MgI  $\lambda 5200$ absorption complex,  at least
blueward  of  $\sim  9200$\AA.   Between  $\sim  9200$\AA\  and  $\sim
9700$\AA, the spectrum is significantly affected by spurious residuals
from the  correction for telluric  absorption, so the spectrum  is not
reliable. At the  redshift of the blazar the MgI  complex would lie at
$9850$\AA, but  again no obvious absorption feature  can be identified
in Figure~\ref{fig1}.

\begin{figure*}
\centerline{ \psfig{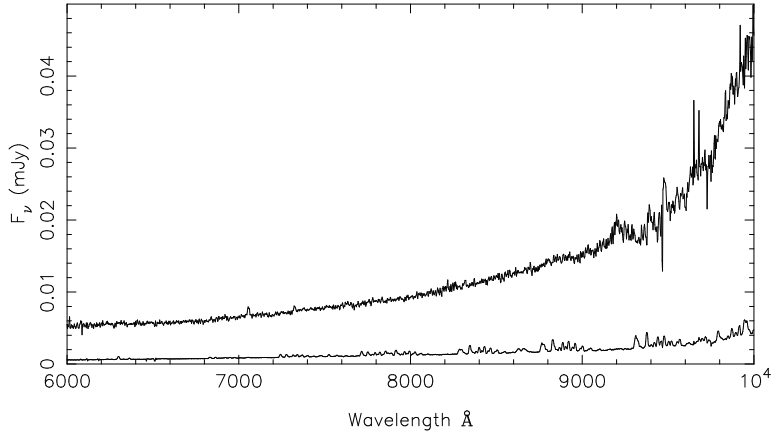} }
\caption[]{The  residual between  the  spectrum of  \pk\ extracted  in
$6\scmd$ and $4\scmd$ bands. Again, no significant absorption features
indicative of a normal galaxy  are apparent. The emission features are
probably  due  to  the  \bl\  source.   The  noise  spectrum  is  also
presented.}
\label{fig2}
\end{figure*}

To further search for any evidence of an absorption system the spectra
from the different  width bands were subtracted; as  the emission from
any  galaxy component  will be  extended, compared  to  the point-like
emission from the quasar source,  we would expect that the residual of
the subtraction between bands of  different width would help to remove
the emission from  the \bl\ in \pk, leaving  proportionally more light
from  the foreground  galaxy.   In Figure~\ref{fig2},  we present  the
difference  between the spectra  extracted from  the $6\scmd$  and the
$4\scmd$  bands; given the  better than  $1\scmd$ seeing  during these
observations, virtually  all of the point-like emission  from the \bl\
source  should   be  contained   within  the  $4\scmd$   band.   Close
examination of this residual  spectrum, however, also reveals a number
of the  emission features identified in  Figure~\ref{fig1}, and, given
the broad width  of the $H_\beta$ emission, these appear  to be due to
the  \bl\  source rather  than  being  galactic  in origin.   As  with
Figure~\ref{fig1}, no significant  absorption features indicative of a
normal  galaxy spectrum  are visible  in Figure~\ref{fig2},  making it
impossible  to determine  the redshift  of the  extended  emission, if
present.

\subsection{HST Observations}
Is  there any  indication of  the  presence of  a galaxy  in the  \pk\
system?  As discussed in  Section~\ref{pks}, the existence of a galaxy
component in this system  has a checkered history, originally detected
in  the imaging  work of  Stickel  \etal\ (1988),  with later  studies
failing  to  identify   any  extended  emission  (Falomo~\etal\  1992;
Kotilainen~\etal\  1998).  To address  this question,  high resolution
images of \pk\ are required.  Such observations were taken on the 29th
of March 1996  with the Planetary Camera chip  of the WFPC2 instrument
on board the Hubble Space Telescope,  as part of a larger survey of BL
Lac hosts  (Urry \etal\  1997; Urry \etal\  1999). Four  exposures, of
duration  14~sec, 80~sec,  and  $2\times 260$~sec,  all  in the  F702W
filter (a  585\AA\ wide passband  centered on 6913\AA),  were obtained
from the HST archive.

\subsubsection{Data Reduction}\label{datareduction}
\begin{figure*}
\centerline{ 
\psfig{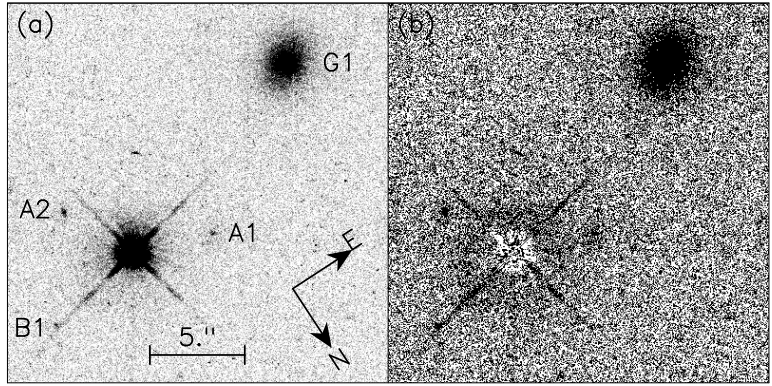} 
}
\caption[]{The left-hand panel presents  the ``raw'' HST image of \pk,
obtained with  WFPC2. \pk\ appears  very point-like. The  companion at
z=0.186 is clearly visible (G1), as well as additional emission in the
vicinity of  the \bl\  source (A1, A2,  \& B1).  The  right-hand panel
presents the  same image with a  modified contrast level  and with the
\bl\ emission  subtracted via the  modeling of the WFPC2  point spread
function; beyond the inner $1\scmd$, some very faint extended emission
can be seen around \pk.}
\label{fig3}
\end{figure*}

A median-combined image of the PC frames is presented in the left-hand
panel of Figure~\ref{fig3}; the  \bl\ appears point-like in this image
and the companion galaxy, at ${\rm z=0.186}$ is clearly visible to the
east of this source (c.f.  Stickel~\etal\ 1988).  Three other `images'
are noted;  two of these,  marked A1 and  A2 have been noted  in other
imaging campaigns (Stickel~\etal\  1988; Falomo~\etal\ 1992), while B1
is noted here for the first time.  The nature of these sources will be
discussed in more detail in Section~\ref{conclusions}.

To search  for extended emission  beneath \pk, a  point-source profile
fit  and subtracted  from the  observed \bl\  profile.  For  this, the
point-spread  function was  modeled  using the  Tiny-Tim algorithm  of
Krist (1995).  The subtracted image is presented in the right-panel of
Figure~\ref{fig3}.  Examination of this  panel does suggest that there
is faint, non-axisymmetric extended emission around the \bl\ source in
\pk.

\begin{figure*}
\vbox{ \centerline{ \psfig{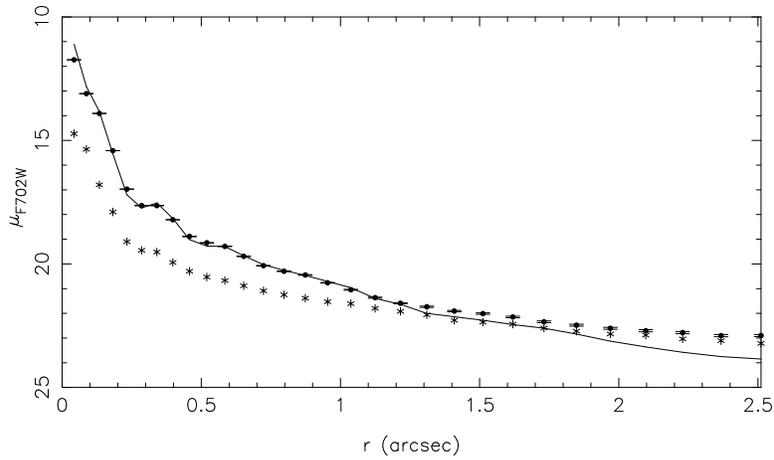} }}
\caption[]{The radial profile of \pk\ (dots) and the galaxy G1 (stars)
as determined  from the WFPC2 image. The  point-spread function (line)
was modeled using the Tiny-Tim algorithm of Krist (1995).}
\label{fig4}
\end{figure*}

To investigate this extended  emission further, the radial profiles of
both  \pk\ and  the companion  galaxy  (G1) where  determined and  are
plotted   together  with   the  Tiny-Tim   point-spread   function  in
Figure~\ref{fig4}.   In the  inner regions  ($\simlt1\scmd$),  \pk\ is
extremely  well   fit  by  the  modeled   point-spread  function.   At
$1\scnd2$,  however, \pk\  begins  to deviate  from  this model,  with
excess flux clearly  seen at larger radii, at a  level very similar to
that of the  companion galaxy (G1); this is  the conclusion reached by
Stickel~\etal\ (1988).  But again,  is this extended emission real? Is
the Tiny-Tim model adequate in describing the point-spread function in
these Planetary Camera frames?

We note that the surface brightness profile presented by Falomo~\etal\
(1992), from  NTT images, is  similarly well described in  the central
regions by  a stellar point-spread function.   However, their Figure~4
shows that there is a systematic enhancement of the surface brightness
profile  with respect to  the point-spread  function, at  radii beyond
$\sim2\scmd$.   Though  they  concluded   that  this  excess  was  not
significant, due to their  larger photometric uncertainties, this same
discrepancy   is  seen   to  begin   at   the  same   radius  in   our
Figure~\ref{fig4},   derived  from  the   ten  times   higher  spatial
resolution  HST data.   This consistency  between our  experiment, and
that of Falomo~\etal\ (1992), indicates  that it is likely that we are
detecting a galactic component in the \pk\ system [this conclusion was
also reached by Stickel Fried \& Kuehr (1993)].

\section{Conclusions}\label{conclusions}
In this paper we have  presented both spectroscopy and imaging of \pk;
this \bl\  source is  seen to possess  extreme brightness  and violent
variability and it has been  suggested that it is being microlensed by
stars in a foreground galaxy.  The apparent detection of such a system
added weight  to this suggestion (Stickel~\etal\  1988), although more
recent observations have cast doubt on that claim (Falomo~\etal\ 1992;
Kotilainen~\etal\ 1998).   The microlensing hypothesis,  however, does
suffer from several  drawbacks and has led to  the suggestion that the
redshift of \pk\ was  misidentified (Narayan \& Schneider 1990); given
the typically  featureless spectrum exhibited  by \bl\ objects  such a
conclusion  is entirely reasonable.   The observations  presented here
appear  to  have  fortuitously caught  \pk\  at  a  low point  in  its
variability  and several  prominent emission  features arising  in the
\bl\ are apparent in the NTT spectra.  These confirm \pk's redshift to
be  $z=0.892\pm0.001$.   At this  redshift,  \pk\  possesses an  ${\rm
M_B\sim-28.5~(\Omega_o=1~\&~h=0.5)}$,  placing  it  amongst  the  most
luminous of quasars.

An  examination  of these  spectra  failed  to  reveal any  absorption
signature  indicative  of the  putative  foreground galaxy  supposedly
responsible for  microlensing the  \bl\ source.  While  several recent
observations have failed  to identify any signature of  such a galaxy,
the HST/WFPC2 imaging presented here apparently supports the existence
of  extended   emission  around  \pk.   Given  the   problems  of  the
microlensing model  in explaining the properties of  \pk, we speculate
that this extended emission arises in the host of the \bl, rather than
in some  intervening galaxy.  The  reality and nature of  the extended
emission can only be investigated with further observations.

Does gravitational  lensing significantly  influence our view  of \pk?
With the lack of any significant  lensing mass along the line of sight
the answer must be no, and considering that the ${\rm z=0.186}$ galaxy
$\sim11\scmd$  from  the  \bl\  can  induce  a  gravitational  lensing
amplification of only a few percent, \pk\ must be appreciated for what
it  is, a luminous  and highly  variable member  of the  active galaxy
family. 

As pointed  out in Section~\ref{datareduction}, an  examination of the
environment  of  \pk\  reveals  the  presence of  several  small,  but
extended  objects;  two of  these  (A1 \&  A2)  are  also apparent  in
previous  images,  including  those  of Stickel~\etal\  (1988).   Such
multiple  components are seen  in another  bright, high  redshift \bl,
AO~0235+164 ($z\sim0.94$); redshift  measurements have determined that
these systems are foreground to  the \bl\ source and represent a group
of galaxies which appear to be responsible for gravitationally lensing
AO~0235+164 (Stickel~\etal\ 1998a).  What is the nature of the systems
near  \pk?  Given  their  small, sub-arcsecond  scale,  they are  most
likely not  members of a group  of massive galaxies along  the line of
sight,  as in  AO~0235+164.   Two possibilities  do, however,  present
themselves:
\begin{itemize}
\item {\bf Environment: }  These systems may represent true companions
of   the   \bl\   source.    At  $z=0.892$,   $1\scmd   \sim8\kpc~{\rm
(\Omega_o=1~\&~h=0.5)}$,  and these  galaxies would  be  $\sim1$kpc in
extent and be  separated from the AGN by  $\sim32$kpc, suggesting that
they could be dwarf satellites of \pk.
\item  {\bf Gravitational  Lensing:  } Could  \pk\ be  gravitationally
lensing  more distant  sources?   The additional  components are  very
similar in  both morphology and  scale to several  recently identified
gravitationally lensed images of high redshift sources identified with
the HST (c.f. Ratnatunga~\etal\  1999).  Considering the separation of
the various components from the  \bl\ and adopting a simple isothermal
mass model to describe  the gravitational lensing, the one-dimensional
velocity dispersion  of this potential would  be $550\kms$, suggesting
that  \pk\ resides  in  a group/small  cluster environment  (Mortlock,
Webster  \&  Hewett  1996).   Again,  further  observations  would  be
required  to confirm  this hypothesis,  but gravitational  lensing may
still play a role in our view of the \pk\ system.
\end{itemize}

\newpage

\newpage

\newpage

\newpage

\end{document}